\documentclass[12pt]{article}
\usepackage{amssymb}
\usepackage{amsmath}

\begin{document}

\title{\hspace{4.1in}\\
\hspace{4.1in}{\small OUTP 0420P}\bigskip \\
$SU(3)$ family symmetry and neutrino bi-tri-maximal mixing}
\author{ Ivo de Medeiros Varzielas and Graham G. Ross \\
The Rudolf Peierls Centre for Theoretical Physics\\
The University of Oxford\\
1 Keble Road \\
Oxford \\
OX1 3NP}
\date{}
\maketitle

\begin{abstract}
The observed large mixing angles in the lepton sector may be the first
signal for the presence of a non-Abelian family symmetry. However, to obtain
the significant differences between the mixing of the neutrino and charged
fermion sectors, the vacuum expectation values involved in the breaking of
such a symmetry in the two sectors must be misaligned. We investigate how
this can be achieved in models with an $SU(3)_{f}$ family symmetry
consistent with an underlying GUT. We show that such misalignment can be
achieved naturally via the see-saw mechanism. We construct a specific example in which the vacuum (mis)alignment is guaranteed by additional
symmetries. This model generates a fermion mass structure consistent with
all quark and lepton masses and mixing angles. Neutrino mixing is close to
bi-tri-maximal mixing.
\end{abstract}

\section{Introduction}

The need to explain the observed pattern of quark and lepton masses and
mixing angles remains a central issue in our attempt to construct a theory
beyond the Standard Model. Perhaps the most conservative possible
explanation is that the symmetry of the Standard Model is extended to
include a Grand Unified and/or family symmetry which order the Yukawa
couplings responsible for the mass matrix structure.

If one restricts the discussion to the quark sector it is possible to build
quite elegant examples involving a spontaneously broken family symmetry
which generates the observed hierarchical structure of masses and mixing
angles. However, attempts to extend this to the leptons has proved very
difficult, mainly because the large mixing angles needed to explain neutrino
oscillation are quite different from the small mixing angles observed in the
quark sector. Indeed the present experimental situation is consistent with
the Harrison, Perkins and Scott ``bi-tri-maximal mixing'' scheme \cite{HPS}
in which the atmospheric neutrino mixing angle is maximal ($sin^{2}(\theta_{@}) = 1/2)$ and the
solar neutrino mixing is tri-maximal ($sin^{2}(\theta_{\odot}) = 1/3)$). If the mixing indeed proves
to be close to bi-tri-maximal mixing it will strongly suggest that the
family symmetry is non-Abelian, because equality between mixing angles
involving different families requires a symmetry relating the magnitude of
the Yukawa couplings of these families, something an Abelian symmetry cannot
do.

In this paper we discuss how bi-tri-maximal mixing can emerge in a theory
with an underlying $SU(3)_{f}$ family symmetry or with a non-Abelian
discrete subgroup of $SU(3)_{f}$. This is of particular interest as $%
SU(3)_{f}$ is the largest family symmetry that commutes with $SO(10)$ and so
fits nicely with promising Grand Unified extensions of the Standard Model.
However, by itself, $SU(3)_{f}$ does not explain why the mixing angles are
small in the quark sector while they are large in the lepton sector. If it
is to be consistent with an underlying spontaneously broken family symmetry
there must be a mismatch between the symmetry breaking pattern in the quark
and charged lepton sectors and the symmetry breaking pattern in the neutrino
sector. In the quark sector and charged lepton sectors the first stage of
family symmetry breaking, $SU(3)_{f}\rightarrow SU(2)_{f},$ generates the
third generation masses while the remaining masses are only generated by the
second stage of breaking of the residual $SU(2)_{f}$. However, in the
neutrino mass sector the dominating breaking must be rotated by $\pi /4$
relative to this, so that an equal combination of $\nu _{\tau }$ and $\nu
_{\mu }$ receives mass at the first stage of mass generation. The subsequent
breaking generating the light masses must also be misaligned by approximately the tri-maximal angle in order to describe solar neutrino oscillation.

There has already been significant progress in constructing models with a
non-Abelian family symmetry capable of generating bi-tri-maximal mixing.
These models are based on a spontaneously broken $A_{4}$ discrete symmetry.
Of course, as just discussed, these models also have to arrange the mismatch
of symmetry breaking in the charged and neutral sectors. In \cite{Ma}, Ma
demonstrated that, assuming such a pattern of symmetry breaking, there is
left an unbroken discrete subgroup of $A_{4}$ which guarantees the
bi-tri-maximal mixing structure. Altarelli and Feruglio \cite{AF} have
recently analysed the scalar potential for a particular $A_{4}$ model and
shown that, at least in a five dimensional version of the model, it is
possible to achieve naturally the desired pattern of symmetry breaking.

It might seem that these $A_{4}$ based models provide examples of the class
of discrete subgroups of the $SU(3)_{f}$ family symmetry discussed above%
\footnote{$A_{4}$ is isomorphic to the dihedral group $\Delta _{12}$ and is
a discrete subgroup of $SO(3)$ as well as $SU(3)$.}. However, in these
models the $A_{4}$ is not a subgroup of the $SU(3)_{f}$ family symmetry
which commutes with $SO(10)$ because the left-handed leptons are assigned to
different $A_{4}$ representations to those assignments for the charge
conjugate of the right-handed leptons. While this is perfectly possible it
does mean that there is no straightforward way to embed these models in a
Grand Unified structure and so some of the attractive features of such a
structure are lost. In particular in the model of \cite{Ma} the quarks
remain massless at the stage the bi-tri-maximal mixing has been generated
for the leptons. The quark masses and mixing angles appear only after
further symmetry breaking and so are completely unrelated to the lepton mass
structure. A possible exception to this approach is \cite{Frigerio}, which does assign all leptons to the same representation of $A_{4}$ and can still generate maximal atmospheric angle.

Given this, we consider it interesting to ask whether bi-tri-maximal mixing
can emerge in a theory with an underlying $SO(10)\otimes SU(3)_{f}$
symmetry \footnote{ See \cite{Maxim} for a review of previous work on $SU(3)$ family symmetry, including original references }
, capable of preserving the phenomeonologically successful GUT
relations between quark and lepton masses. The model is heavily constrained
because the $SU(3)_{f}$ multiplet assignment of all the quarks and leptons
must all be the same. Nevertheless we show that it is possible to build a
model capable of describing all quark and lepton masses and mixing angles in
which bi-tri-maximal mixing emerges naturally. In this model there is a
close relation between quark and lepton masses and the Georgi-Jarlskog
relation between charged lepton and quark masses \cite{GJ} is readily
obtained. The symmetric structure of the mass matrices by the multiplet
assignments also allows it to reproduce the phenomenologically successful
Gatto-Sartori-Tonin (GST) relation \cite{GST} for the $(1,2)$ sector
mixing.

While completing this paper we received a paper by King \cite{SO3} who shows
how to achieve bi-tri-maximal structure using $SO(3)$ as the family
symmetry. The model is similar in general structure to the one presented
here both implementing the general strategy explored in \cite{SU3},\cite%
{Vives}. The main difference, apart from the different choice of family
symmetry group, is that the model of \cite{SO3} assigns left-handed states
to different $SO(3)$ representations from those of the charge conjugate of
the right-handed states, generating non-symmetric mass matrices. As a result
the GST relation is lost and the model does not straightforwardly extend to
an underlying $SO(10)$ unification.

As discussed above one of the main difficulties in realising bi-tri-maximal
mixing in a $SU(3)_{f}$ family symmetry model is the need to explain why the
dominant breaking leading to the generation of third generation masses in
the quark sector is not the dominant effect in the neutrino sector too. At
first sight it appears quite unnatural. However, if neutrino masses are
generated by the see-saw mechanism \cite{Seesaw} in fact it can readily
arise even if all quark and lepton Dirac mass matrices, including those of
the neutrinos, have similar forms up to Georgi-Jarlskog type factors. To see
this consider the general form of the see-saw mechanism 
\begin{equation*}
\;M_{\nu }=\quad M_{D}^{\nu }\;M_{M}^{-1}\;M_{D}^{\nu T}
\end{equation*}%
where $M_{\nu }$ is the mass matrix for the light neutrino states, $%
M_{D}^{\nu }$ is the Dirac mass matrix coupling $\nu $ to $\nu ^{c}$ and $%
M_{M}$ is the Majorana mass matrix coupling $\nu ^{c}$ to $\nu ^{c}$. We
consider the case where the Majorana mass matrix also has an hierarchical
structure of the form 
\begin{equation*}
M_{M}\simeq \left( {{\begin{array}{*{20}c} {M_1 } \hfill & \hfill & \hfill
\\ \hfill & {M_2 } \hfill & \hfill \\ \hfill & \hfill & {M_3 } \hfill \\
\end{array}}}\right) \quad M_{1}<<M_{2}<<M_{3}.
\end{equation*}%
For a sufficiently strong hierarchy this gives rise to sequential domination 
\cite{SeqDom} in which the heaviest of the three light eigenstates gets its
mass from the exchange of the lightest right-handed singlet neutrino with
mass $M_{1}$. In this case the contribution to the light neutrino mass
matrix of the field responsible for the dominant $(3,3)$ terms in the Dirac
mass matrices is suppressed by the relative factor $M_{1}/M_{3}$ and may
readily be subdominant in the neutrino sector. The message from this is that
any underlying quark-lepton symmetry is necessarily broken in the neutrino
sector due to the Majorana masses of the right-handed neutrino states and,
through the see-saw mechanism, this feeds into the neutrino masses and the
lepton mixing angles. This example illustrates how this effect can hide an
underlying quark-lepton symmetry in the Dirac mass sector.

In this paper we implement this structure to build a model with near
bi-tri-maximal mixing. We consider only the case of a supersymmetric
extension of the Standard Model because only in these models is the
hierarchy problem associated with a high-scale GUT under control. Rather
than work with a complete $SO(10)\otimes SU(3)_{f}$ theory (which, in a
string theory, may only be relevant above the string scale) we consider here
the case where the gauge symmetry is $G_{PS}\otimes SU(3)_{f}$ \ where $%
G_{PS}$ is the Pati-Salam group $G_{PS}\equiv SU(4)_{PS}\otimes
SU(2)_{L}\otimes SU(2)_{R}$. The $SU(3)_{f}$ representation assignments are
chosen in a way consistent with this being a subgroup of $SO(10)\otimes
SU(3)_{f}$. The construction of the model closely follows that of \cite{SU3}
and \cite{Vives}, and proceeds by identifying a simple $U(1)\otimes
U^{\prime }(1)$ symmetry capable of restricting the allowed Yukawa couplings
to give viable mass matrices for both the quarks and leptons. We pay
particular attention to an analysis of the scalar potential which is
responsible for the vacuum alignment generating bi-tri-maximal mixing.

The Majorana mass matrices are generated by the lepton number violating
sector, and we find it to be helpful that the dominant contribution to the
Majorana mass matrix for the neutrinos is aligned along the $3$rd direction,
as is the case for the fermion Dirac matrix. The major difference is the
ordering of the hierarchical structure in the two sectors. As we will show,
in this case it is possible to achieve bi-tri-maximal mixing very closely,
with deviations coming from the charged lepton sector. This type of situation is described in some detail in \cite{Rode}.

The organisation of the paper is as follows. In Section \ref{sec:Symmetries}
we introduce the symmetry assignment of the states of the Standard Model
together with the fields needed to implement a viable pattern of symmetry
breaking. We discuss how this symmetry breaking leads to an effective low
energy theory of fermion masses through the Froggatt-Nielsen mechanism \cite{SU3} \cite{Froggatt:1978nt}. In Section \ref{sec:Masses} we list the
dominant Yukawa couplings in the effective theory responsible for fermion
masses and compute the Dirac and Majorana mass matrices. The details of the
symmetry breaking alignment are presented in the Appendix where we discuss
the details of the minimisation of the effective potential, particularly the
minimisation\ and effect of the $D$-terms and the effect of the soft SUSY
breaking masses. In Section \ref{sec:Phenomenology} we discuss the
phenomenological implications of this model, and estimate the magnitude of
the corrections to bi-tri-maximal mixing. Finally in Section \ref%
{sec:Conclusion} we present the conclusions.

\section{The construction of $SU(3)_{f}$ family symmetry models \label{sec:Symmetries}}

\subsection{Symmetries}

As discussed above we will start with the gauge group $G_{PS}\otimes
SU(3)_{f}$. We wish to assign our states to representations in a manner
consistent with an underlying $SO(10)\otimes SU(3)_{f}$ symmetry so we will
discuss the representation content as if this is the gauge group even though
we will use only the $SU(4)_{PS}\otimes SU(2)_{L}\otimes SU(2)_{R}\otimes
SU(3)_{f}$ subgroup in constructing models. The Standard Model (SM) fermions 
$\psi _{i},\psi _{j}^{c}$ \footnote{$i,j=1,2,3$ are family indices} are
assigned to a $(\mathbf{16},\mathbf{3})$ representation of $SO(10)\otimes
SU(3)_{f}$. The Higgs doublets of the SM are part of a $(\mathbf{10},\mathbf{%
1})$ representation, $H$. In addition we introduce an adjoint of Higgs
fields $H_{45}\sim (\mathbf{45},\mathbf{1})$, which in our effective theory
has a vacuum expectation value (vev) consistent with the residual $G_{PS}\otimes SU(3)_{f}$ symmetry which leaves the hypercharge $%
Y=T_{3R}+(B-L)/2$ unbroken \cite{SU3}.

To recover the SM we must completely break the family symmetry. We will do
so in steps, first with a dominant breaking from $SU(3)_{f}\rightarrow
SU(2)_{f}$ and then the breaking of the remaining $SU(2)_{f}$. This
spontaneous symmetry breaking will be achieved by additional Higgs fields
that are either triplets ($\mathbf{3}_{i}$) or anti-triplets ($\bar{\mathbf{3%
}}^{i}$) of the family $SU(3)_{f}$, and the alignment of their vacuum
expectation values (vevs) is a principal concern of this paper. In a
realistic model it is necessary to extend the symmetry in order to eliminate
terms in the effective lagrangian. The construction of a specific model
requires that the full multiplet content is specified together with its
transformation properties under $G_{PS}\otimes SU(3)_{f}$ and under the
addition symmetry needed to limit the allowed couplings. In the model
constructed here the the additional symmetry is $U(1)\otimes U^{\prime }(1)$%
. The multiplet content and transformation properties for the model are
given in Table \ref{Ta:Table 1}. In addition to the fields discussed above
it includes the fields $\theta $ and $\bar{\theta }$ whose vevs break $%
SU(4)_{PS}$ and lepton number and generate the Majorana mass matrix. There
are also additional $SU(4)_{PS}\otimes SU(2)_{L}\otimes SU(2)_{R}$ singlet
fields needed for vacuum alignment as discussed in the Appendix.

\bigskip

\begin{table}[tbp] \centering%

\begin{tabular}{|c||c||c|c|c||c|c|c|}
\hline
Field & $SU(3)_{f}$ & $SU(4)_{PS}$ & $SU(2)_{L}$ & $SU(2)_{R}$ & $R$ & $U(1)$
& $U(1)^{ \prime }$ \\ \hline\hline
$\psi $ & $\mathbf{3}$ & $\mathbf{4}$ & $\mathbf{2}$ & $\mathbf{1}$ & $%
\mathbf{1}$ & $\mathbf{0}$ & $\mathbf{0}$ \\ 
$\psi ^{c}$ & $\mathbf{3}$ & $\bar{\mathbf{4}}$ & $\mathbf{1}$ & $\mathbf{2}$
& $\mathbf{1}$ & $\mathbf{0}$ & $\mathbf{0}$ \\ \hline\hline
$\theta$ & $\mathbf{3}$ & $\bar{\mathbf{4}}$ & $\mathbf{1}$ & $\mathbf{2%
}$ & $\mathbf{0}$ & $\mathbf{0}$ & $\mathbf{0}$ \\ 
$\bar{\theta}$ & $\bar{\mathbf{3}}$ & $\mathbf{4}$ & $\mathbf{1}$ & $%
\mathbf{2}$ & $\mathbf{0}$ & $\mathbf{0}$ & $\mathbf{0}$ \\ \hline\hline
$H$ & $\mathbf{1}$ & $\mathbf{1}$ & $\mathbf{2}$ & $\mathbf{2}$ & $\mathbf{0}
$ & $-\mathbf{4}$ & $-\mathbf{4}$ \\ 
$H_{45}$ & $\mathbf{1}$ & $\mathbf{15}$ & $\mathbf{1}$ & $\mathbf{3}$ & $%
\mathbf{0}$ & $\mathbf{2}$ & $\mathbf{2}$ \\ \hline\hline
$\phi_{3}$ & $\mathbf{3}$ & $\mathbf{1}$ & $\mathbf{1}$ & $\mathbf{1}$ & $%
\mathbf{0}$ & $-\mathbf{2}$ & $-\mathbf{3}$ \\ 
$\bar{\phi}_{3}$ & $\mathbf{\bar{3}}$ & $\mathbf{1}$ & $\mathbf{1}$ & $%
\mathbf{3}\oplus \mathbf{1}$ & $\mathbf{0}$ & $\mathbf{2}$ & $\mathbf{2}$ \\ 
$\phi_{2}$ & $\mathbf{3}$ & $\mathbf{1}$ & $\mathbf{1}$ & $\mathbf{1}$ & $%
\mathbf{0}$ & $-\mathbf{1}$ & $\mathbf{1}$ \\ 
$\bar{\phi}_{2}$ & $\bar{\mathbf{3}}$ & $\mathbf{1}$ & $\mathbf{1}$ & $\mathbf{1}$
& $\mathbf{0}$ & $-\mathbf{1}$ & $\mathbf{1}$ \\ 
$\phi_{23}$ & $\mathbf{3}$ & $\mathbf{1}$ & $\mathbf{1}$ & $\mathbf{1}$
& $\mathbf{0}$ & $-\mathbf{4}$ & $-\mathbf{3}$ \\ 
$\bar{\phi}_{23}$ & $\bar{\mathbf{3}}$ & $\mathbf{1}$ & $\mathbf{1}$ & $%
\mathbf{1}$ & $\mathbf{0}$ & $\mathbf{1}$ & $\mathbf{1}$ \\ 
$\phi_{123}$ & $\mathbf{3}$ & $\mathbf{1}$ & $\mathbf{1}$ & $\mathbf{1}$ & $%
\mathbf{0}$ & $\mathbf{0}$ & $\mathbf{1}$ \\ 
$\bar{\phi}_{123}$ & $\mathbf{\bar{3}}$ & $\mathbf{1}$ & $\mathbf{1}$ & $%
\mathbf{1}$ & $\mathbf{0}$ & $\mathbf{3}$ & $\mathbf{3}$ \\ \hline\hline
$X_{3}$ & $\mathbf{1}$ & $\mathbf{1}$ & $\mathbf{1}$ & $\mathbf{1}$ & $%
\mathbf{2}$ & $\mathbf{0}$ & $\mathbf{1}$ \\ 
$Y_{2}$ & $\mathbf{1}$ & $\mathbf{1}$ & $\mathbf{1}$ & $\mathbf{1}$ & $%
\mathbf{2}$ & $-\mathbf{1}$ & $-\mathbf{3}$ \\ 
$X_{23}$ & $\mathbf{1}$ & $\mathbf{1}$ & $\mathbf{1}$ & $\mathbf{1}$ & $%
\mathbf{2}$ & $\mathbf{1}$ & $\mathbf{0}$ \\ 
$Y_{23}$ & $\mathbf{1}$ & $\mathbf{1}$ & $\mathbf{1}$ & $\mathbf{1}$ & $%
\mathbf{2}$ & $\mathbf{3}$ & $\mathbf{2}$ \\ 
$X_{123}$ & $\mathbf{1}$ & $\mathbf{1}$ & $\mathbf{1}$ & $\mathbf{1}$ & $%
\mathbf{2}$ & $-\mathbf{3}$ & $-\mathbf{4}$ \\ 
$Y_{123}$ & $\mathbf{1}$ & $\mathbf{1}$ & $\mathbf{1}$ & $\mathbf{1}$ & $%
\mathbf{2}$ & $-\mathbf{1}$ & $-\mathbf{2}$ \\ 
$Z_{123}$ & $\mathbf{1}$ & $\mathbf{1}$ & $\mathbf{1}$ & $\mathbf{1}$ & $%
\mathbf{4/3}$ & $-\mathbf{3}$ & $-\mathbf{4}$ \\ \hline\hline
$S_{3}$ & $\mathbf{1} $ & $\mathbf{1}$ & $\mathbf{1}
$ & $\mathbf{1}$ & $ \mathbf{0} $ & $\mathbf{0}$ & $ -\mathbf{1}$ \\
$\Sigma$ & $\mathbf{3} \otimes \bar{\mathbf{3}}$ & $\mathbf{1}$ & $\mathbf{1}
$ & $\mathbf{1}$ & $\mathbf{2/3}$ & $\mathbf{0}$ & $\mathbf{0}$ \\ \hline
\end{tabular}
\caption{Model field and representation content} \label{Ta:Table 1}%
\end{table}%

\bigskip

\subsection{Spontaneous symmetry breaking \label{sec:SSB}}

We now summarise the pattern of family symmetry breaking in our model. The
detailed minimisation of the effective potential which gives this structure
is given in the Appendix.

The dominant breaking of $SU(3)_{f}$ responsible for the third generation
quark and charged lepton masses is provided by the $\bar{\phi}_{3}$ vev 
\begin{equation}
\left\langle \bar{\phi}_{3}\right\rangle =\left( 
\begin{array}{ccc}
0 & 0 & 1%
\end{array}%
\right) \otimes \left( 
\begin{array}{cc}
a_{u} & 0 \\ 
0 & a_{d}%
\end{array}%
\right)  \label{eq:P3B vev}
\end{equation}%
where the $SU(3)\times SU(2)_{R}$ structure is exhibited. To preserve $D$%
-flatness, another field, $\phi _{3}$, also acquires a large vev 
\begin{equation}
\left\langle \phi _{3}\right\rangle \simeq \left( 
\begin{array}{c}
0 \\ 
\delta \\ 
\sqrt{a_{u}^{2}+a_{d}^{2}}%
\end{array}%
\right)  \label{eq:P3 vev}
\end{equation}%
where $\delta $ is a small correction needed to maintain $D$-flatness (%
\textit{c.f.} the Appendix), appearing once the remaining symmetry breaking
is taken into account. Notice that $\bar{\phi}_{3}$ also
breaks the $SU(2)_{R}$ so the residual symmetry is now $SU(4)_{PS}\otimes
SU(2)_{L}\otimes U(1)_{R}.$

The breaking of the remaining family symmetry is achieved when a triplet $%
\phi _{2}$ acquires the vev%
\begin{equation}
\left\langle \phi _{2}\right\rangle \simeq \left( 
\begin{array}{c}
0 \\ 
y \\ 
0%
\end{array}%
\right)  \label{eq:P2 vev}
\end{equation}
Due to the allowed couplings in the superpotential (\textit{c.f.} the
discussion in the Appendix) this vev is orthogonal to $\left\langle \bar{\phi%
}_{3}\right\rangle$.

Further fields acquire vevs constrained by the allowed couplings in the
theory. As detailed in the Appendix the field $\bar{\phi}_{23}$ acquires a
vev
\begin{equation}
\left\langle \bar{\phi}_{23}\right\rangle \simeq \left( 
\begin{array}{ccc}
0 & b & -b%
\end{array}%
\right)  \label{eq:P23B vev}
\end{equation}

Note that it is the underlying $SU(3)_{f}$ that forces the vevs in the $2$nd
and the $3$rd directions to be equal in magnitude, so that the $\bar{\phi}_{23}$ is
rotated by $\pi /4$ relative to the $\bar{\phi}_{3}$ vev. This will be
important in generating an acceptable pattern for quark masses and in
generating bi-maximal mixing in the lepton sector. Finally the fields $\bar{\phi}_{123}$ and $\phi
_{123}$ acquire the vevs
\begin{equation}
\left\langle \bar{\phi}_{123}\right\rangle =\left( 
\begin{array}{ccc}
\bar{c} & \bar{c} & \bar{c}%
\end{array}%
\right) .  \label{eq:P123B vev}
\end{equation}%

\begin{equation}
\left\langle \phi_{123}\right\rangle =\left(%
\begin{array}{c}
c \\ 
c \\ 
c%
\end{array}%
\right)  \label{eq:P123 vev}
\end{equation}
where $c=\bar{c}e^{i\gamma }$.

Its important to note again that, even though $SU(3)_{f}$ is spontaneously
broken by these vevs, it is also responsible for aligning them so that the
elements have equal magnitude. As we shall see this structure is crucial in
obtaining tri-maximal mixing in the solar neutrino sector.

\section{The fermion mass sector \label{sec:Masses}}

\subsection{The effective superpotential}

Having specified the multiplet content and the symmetry properties it is now
straightforward to write down all terms in the superpotential allowed by the
symmetries of the theory. In this Section we concentrate on those terms
responsible for generating the fermion mass matrix. Since we are working
with an effective field theory in which the heavy modes associated with the
various stages of symmetry breaking have been integrated out we must include
terms of arbitrary dimension. In practice it is only necessary to keep the
leading terms needed to give all quarks and leptons a mass and to determine
their mixing angles. For the generation of quark, charged lepton and
neutrino Dirac masses these are 
\begin{equation}
P_{Y}=\frac{1}{M^{2}}\bar{\phi}_{3}^{i}\psi _{i}\bar{\phi}_{3}^{j}\psi
_{j}^{c}H  \label{Y33}
\end{equation}%
\begin{equation}
+\frac{1}{M^{3}}\bar{\phi}_{23}^{i}\psi _{i}\bar{\phi}_{23}^{j}\psi
_{j}^{c}HH_{45}  \label{eq:Y_P23_P23}
\end{equation}

\begin{equation}
+\frac{1}{M^{2}}\bar{\phi}_{23}^{i}\psi _{i}\bar{\phi}_{123}^{j}\psi
_{j}^{c}H  \label{eq:Y_P23_P123}
\end{equation}%
\begin{equation}
+\frac{1}{M^{2}}\bar{\phi}_{123}^{i}\psi _{i}\bar{\phi}_{23}^{j}\psi
_{j}^{c}H  \label{eq:Y_P123_P23}
\end{equation}%

\begin{equation}
+\frac{1}{M^{7}}\bar{\phi}_{2}^{i}\psi _{i}\bar{\phi}_{123}^{j}\psi
_{j}^{c}HH_{45} (\bar{\phi}_{3}^{k} \phi_{3_{k}})^2 \label{eq:Y_P2_P123}
\end{equation}

Note that the terms appear suppressed by inverse powers of a mass scale
which we have generically denoted by $M$. Identification of this scale for
each term is important in studying the phenomenology and to do this one has
to identify how these non-renormalisable terms arise. This occurs at the
stage where superheavy fields, the messenger fields, are integrated out. In
Section \ref{sec:Messengers} we consider this in detail.

The terms allowed by the symmetries responsible for the Majorana mass matrix
involve the $\bar{\theta}^{i}$ anti-triplet responsible for breaking lepton
number and $SU(4)_{PS}$. Its vev is aligned along the 3 direction (%
\textit{c.f.} the discussion in the Appendix). The leading terms are

\begin{equation}
P_{M}=\frac{1}{M}\bar{\theta}^{i}\psi _{i}^{c}\bar{\theta}^{j}\psi _{j}^{c}
\label{eq:M_th_th}
\end{equation}%
\begin{equation}
+\frac{1}{M^{5}}\bar{\phi}_{23}^{i}\psi _{i}^{c}\bar{\phi}_{23}^{j}\psi
_{j}^{c}\bar{\theta}^{k}\phi _{123_{k}}\bar{\theta}^{l}\phi _{3_{l}}+\frac{1%
}{M^{5}}\bar{\theta}^{i}\psi _{i}^{c}\bar{\phi}_{23}^{j}\psi _{j}^{c}\bar{%
\theta}^{k}\phi _{123_{k}}\bar{\phi}_{23}^{l}\phi _{3_{l}}
\label{eq:M_P23_P23}
\end{equation}%
\begin{equation}
+\frac{1}{M^{5}}\bar{\phi}_{123}^{i}\psi _{i}^{c}\bar{\phi}_{123}^{j}\psi
_{j}^{c}\bar{\theta}^{k}\phi _{23_{k}}\bar{\theta}^{l}\phi _{3_{l}}+\frac{1}{%
M^{5}}\bar{\theta}^{i}\psi _{i}^{c}\bar{\phi}_{123}^{j}\psi _{j}^{c}\bar{\phi%
}_{123}^{k}\phi _{23_{k}}\bar{\theta}^{l}\phi _{3_{l}}+\frac{1}{M^{5}}\bar{%
\theta}^{i}\psi _{i}^{c}\bar{\phi}_{123}^{j}\psi _{j}^{c}\bar{\theta}%
^{k}\phi _{23_{k}}\bar{\phi}_{123}^{l}\phi _{3_{l}}  \label{eq:M_P123_P123}
\end{equation}%
In all these equations we have omitted the overall coupling associated with
each term. These are not determined by the symmetries alone and are all
expected to be of $O(1)$.

\subsection{The messenger sector\label{sec:Messengers}}

The scale $M$ entering in the effective superpotential is set by the heavy
messenger states in the tree diagrams giving rise to the higher dimension
terms. There are two classes of diagram, corresponding to heavy messenger
states transforming as $\mathbf{4}$s under $SU(4)_{PS}$ (vectorlike families) or
those that don't (heavy Higgs). Which class dominates depends on the massive
multiplet spectrum which in turn is specified by the details of the theory
at the high scale. Here we assume that the heavy vectorlike families are the
lightest and dominate.

These states carry the same quantum numbers as one of the external quark or
lepton fields. Due to the underlying $SU(2)_{L}$, the $M_{Q_{L}}$ (the
left-handed quark messenger mass scale) will be the same for the up and down
quarks. However $SU(2)_{R}$ is broken and so the messenger mass $M_{u_{R}}$ 
(the right-handed up quark messenger mass scale) needs not be the same as $%
M_{d_{R}}$ (the right-handed down quark messenger mass scale). The lepton
messenger mass scales have a similar structure, with $M_{L_{L}}$ (the
left-handed lepton messenger mass scale) contributing equally to the charged
lepton and neutrino Dirac couplings, but with $M_{e_{R},\nu _{R}}$ (the
right-handed charged lepton and neutrino messenger mass scale) having
different scales due to $SU(2)_{R}$ breaking effects.

This splitting of the masses of the messengers is important because it is
responsible for the differences between the up and down quark and lepton
masses. As we noted above, the underlying $SO(10)\otimes SU(3)_{f}\otimes
U(1)\otimes U^{\prime }(1)$ structure forces all matter states to have the
same family charges and so the leading terms in the superpotential
contribute equally to all sectors. However the soft messenger masses which
enter the effective lagrangian are sensitive in leading order to $SO(10)$
breaking effects and thus can differentiate between these sectors by fixing
different expansion parameters in the different sectors.

To see what choice for the messenger masses is necessary phenomenologically
we note that a fit to the up and down quark mass matrices has the form \cite{Roberts}

\begin{equation}
Y_{u}\propto \left[ 
\begin{array}{ccc}
0 & \epsilon _{u}^{3} & {O}\left( \epsilon _{u}^{3}\right) \\ 
. & \epsilon _{u}^{2} & {O}\left( \epsilon _{u}^{2}\right) \\ 
. & . & 1%
\end{array}%
\right]  \label{eq:Yu}
\end{equation}

\begin{equation}
Y_{d}\propto \left[ 
\begin{array}{ccc}
0 & 1.5\epsilon _{d}^{3} & 0.4\epsilon _{d}^{3} \\ 
. & \epsilon _{d}^{2} & 1.3\epsilon _{d}^{2} \\ 
. & . & 1%
\end{array}%
\right]  \label{eq:Yd}
\end{equation}%
with expansion parameters 
\begin{equation}
\epsilon _{u}\simeq 0.05,\epsilon _{d}\simeq 0.15  \label{eq:u,d eps}
\end{equation}

From eq(\ref{eq:M_P23_P23}) it may be seen that in the quark sector the
expansion parameters in the $(2,3)$ block are essentially determined by the $%
\bar{\phi}_{23}$ vev divided by the relevant messenger mass scale. If the
expansion parameters are to differ it is necessary for $M_{Q_{L}}$ to be
larger than the other messenger masses in which case 
\begin{equation}
\epsilon _{u,d}\simeq \frac{b}{M_{u_{R},d_{R}}}
\end{equation}%
Clearly to generate the form of eq(\ref{eq:u,d eps}) we then need%
\begin{equation}
M_{d_{R}}\simeq \frac{1}{3}M_{u_{R}}\ll M_{Q_{L}}
\end{equation}

In the lepton sector we know that the $SU(5)$ relation $m_{b}\simeq m_{\tau
} $ at the unification scale is, after including radiative corrections, in
good agreement with the measured masses. For this to be the case here it is
necessary that the breaking of the $SU(4)_{PS\text{ }}$ should not dominate
the contribution to the down sector messenger masses. This is to be expected
in these models because the breaking of $SU(2)_{R}$ occurs through the lepton
number breaking sector which does not couple in leading order to the the
right-handed charged lepton messenger states. With 
\begin{eqnarray*}
M_{e_{R}} &\simeq &M_{d_{R}} \\
M_{Q_{L}} &\simeq &M_{L_{L}}
\end{eqnarray*}%
the right-handed messengers dominate and the relation $m_{b}\simeq m_{\tau }$
follows. However the right-handed neutrino messengers do couple in leading
order to the $SU(2)_{R}$ breaking fields and so may be expected to be
anomalously heavy. This is helpful, because a small right-handed neutrino
expansion parameter $\epsilon _{\nu _{R}}$ naturally explains the large
hierarchical structure of Majorana masses which, as discussed above, is
needed to overcome the large Dirac neutrino mass in the $(3,3)$ direction.

The expansion parameters in the lepton sector are then given by%
\begin{equation}
\epsilon _{\nu _{R},\nu _{L},e_{R}}\simeq \frac{b}{M_{\nu _{R},L_{L},e_{R}}}
\label{eq:L eps}
\end{equation}

Bounds on the messenger masses of the neutrinos (or rather, the associated
expansion parameters) will be presented in Section \ref{sec:Phenomenology}.
The other expansion parameters are chosen to fit the masses as in eq(\ref%
{eq:u,d eps}).

Note that the contribution to the $(3,3)$ entries of the quark and charged
lepton mass matrices involves the combination $a_{u}/M_{u_{R}}$ and $%
a_{d}/M_{d_{R}}=a_{d}/M_{e_{R}}$ for the up and down sectors respectively.
If $\bar{\phi}_{3}$ dominates the ($SU(2)_{R}$ breaking) contribution to the
messenger masses then we expect $a_{u}/M_{u_{R}}\simeq
a_{d}/M_{d_{R}}\simeq 1$ which is indeed the phenomenologically desirable
choice \cite{SU3}.

\subsection{The Dirac mass matrix structure \label{sec:Dirac}}

Using the expansion parameters introduced above we can now write the
approximate quark mass matrices for the second and third generations
following from eqs(\ref{Y33}) and (\ref{eq:Y_P23_P23}) in the form 
\begin{equation}
Y_{u}\propto \left( 
\begin{array}{cc}
-2\epsilon _{u}^{2} \frac{\epsilon_{u}}{\epsilon_{d}} & 2\epsilon _{u}^{2} 
\frac{\epsilon_{u}}{\epsilon_{d}} \\ 
2\epsilon _{u}^{2} \frac{\epsilon_{u}}{\epsilon_{d}} & 1%
\end{array}%
\right) ,Y_{d}\propto \left( 
\begin{array}{cc}
\epsilon _{d}^{2} & -\epsilon _{d}^{2} \\ 
-\epsilon _{d}^{2} & 1%
\end{array}%
\right)  \label{eq:Y_Q_23 matrices}
\end{equation}

In writing this form we have made a choice for the $H_{45}$ vev that appears
in the terms contributing to these elements. $\left\langle H_{45} \right\rangle$ preserves $%
G_{PS}$ and is proportional to the hypercharge $Y=T_{3R}+(B-L)/2$ \cite{SU3}, \cite{liliana}. To fit the strange quark mass we take its magnitude to
be such that 
\begin{equation*}
\frac{ \left\langle H_{45} \right\rangle}{M}|_{d}\equiv \frac{Y\left( d^{c}\right) h_{45}}{M_{d_{R}}}%
\approx O(1)
\end{equation*}%
and with this choice, the factor $Y\left( u_{R}\right) /Y\left( d_{R}\right) =-2$
appears in $Y_{u}.$

Because the charged lepton messengers have the same messenger mass scale as
the down quarks, the charged lepton mass matrix is similar to $Y_{d}$ with
the form 
\begin{equation}
Y_{l}\propto \left( 
\begin{array}{cc}
3\epsilon _{d}^{2} & -3\epsilon _{d}^{2} \\ 
-3\epsilon _{d}^{2} & 1%
\end{array}%
\right)  \label{eq:Y_c_23 matrix}
\end{equation}%
where the Georgi-Jarlskog factor $Y\left( e_{R}\right) /Y\left( d_{R}\right)
=3$ comes from  $\left\langle H_{45} \right\rangle$. This factor gives $m_{\mu
}\simeq 3m_{s}$ which is, after including radiative corrections, in good
agreement with the measured masses, an obvious advantage to having an
underlying GUT.

Having explained the origin of the structure in the $(2,3)$ block it is
straightforward to follow the origin of the full three generation Yukawa
matrices for the quarks and leptons. Including the effect of the terms in
eqs(\ref{eq:Y_P23_P123}) and (\ref{eq:Y_P123_P23}), we have%
\begin{equation}
Y_{u}\propto \left( 
\begin{array}{ccc}
0 & g_{\odot }\epsilon _{u}^{2}\epsilon _{d} & -g_{\odot }\epsilon
_{u}^{2}\epsilon _{d} \\ 
g_{@}\epsilon _{u}^{2}\epsilon _{d} & -2\epsilon _{u}^{2}\frac{\epsilon _{u}%
}{\epsilon _{d}} & 2\epsilon _{u}^{2}\frac{\epsilon _{u}}{\epsilon _{d}} \\ 
-g_{@}\epsilon _{u}^{2}\epsilon _{d} & 2\epsilon _{u}^{2}\frac{\epsilon _{u}%
}{\epsilon _{d}} & 1%
\end{array}%
\right)  \label{eq:Y_u matrix}
\end{equation}%
\begin{equation}
Y_{d}\propto \left( 
\begin{array}{ccc}
0 & g_{\odot }\epsilon _{d}^{3} & -g_{\odot }\epsilon _{d}^{3} \\ 
g_{@}\epsilon _{d}^{3} & \epsilon _{d}^{2} & -\epsilon _{d}^{2} \\ 
-g_{@}\epsilon _{d}^{3} & -\epsilon _{d}^{2} & 1%
\end{array}%
\right)  \label{eq:Y_d matrix}
\end{equation}%
\begin{equation}
Y_{l}\propto \left( 
\begin{array}{ccc}
0 & g_{\odot }\epsilon _{d}^{3} & -g_{\odot }\epsilon _{d}^{3} \\ 
g_{@}\epsilon _{d}^{3} & 3\epsilon _{d}^{2} & -3\epsilon _{d}^{2} \\ 
-g_{@}\epsilon _{d}^{3} & -3\epsilon _{d}^{2} & 1%
\end{array}%
\right)  \label{eq:Y_c matrix}
\end{equation}

\begin{equation}
Y_{\nu }\propto \left( 
\begin{array}{ccc}
0 & g_{\odot }\epsilon _{\nu _{L}}^{2}\epsilon _{d} & -g_{\odot }\epsilon
_{\nu _{L}}^{2}\epsilon _{d} \\ 
g_{@}\epsilon _{\nu _{L}}^{2}\epsilon _{d} & (g_{@}+g_{\odot })\epsilon
_{\nu _{L}}^{2}\epsilon _{d} & (g_{@}-g_{\odot })\epsilon _{\nu
_{L}}^{2}\epsilon _{d} \\ 
-g_{@}\epsilon _{\nu _{L}}^{2}\epsilon _{d} & (-g_{@}+g_{\odot })\epsilon
_{\nu _{L}}^{2}\epsilon _{d} & \frac{\epsilon _{\nu _{L}}^{2}}{\epsilon
_{d}^{2}}%
\end{array}%
\right)  \label{eq:Y_nu matrix}
\end{equation}%
In this we have restored the dependence on the unknown Yukawa couplings of $%
O(1)$, after a suitable $O(1)$ redefinition of the expansion parameters. As
will be discussed, we will concentrate on the case $g_{\odot }=g_{@}$, which
is needed for the GST relation \cite{GST}.

The structure of the Dirac neutrino mass matrix $Y_{\nu }$ follows from the
terms in eqs(\ref{eq:Y_P23_P123}) and (\ref{eq:Y_P123_P23}). The form shown
applies in the limit where the dominant carriers are left-handed. As discussed
above this is to be expected because the $SU(2)_{R}$ breaking gives rise to
a very heavy right-handed neutrino messenger mass, corresponding to $\epsilon
_{\nu _{R}}\ll \epsilon _{\nu _{L}}$. The term involving the $H_{45}$ field
in eq(\ref{eq:Y_P23_P23}) is suppressed by an additional messenger mass. For 
$\frac{3\epsilon _{\nu _{L}}^{3}}{2\epsilon _{d}}\ll \epsilon_{\nu_{L}}^{2} \epsilon_{d}$ it is subdominant
leaving the leading terms shown in eq(\ref{eq:Y_nu matrix}). This
corresponds to the upper bound $\epsilon _{\nu _{L}}\ll \frac{2}{3}\epsilon _{d}^{2}$.

Note that the differences between the $(1,2)$ and $(1,3)$ elements of $Y_{d}$,
needed to fit the data c.f. eq.(\ref{eq:Yd}), come from eq(\ref{eq:Y_P2_P123}). 
Thus, due to $H_{45}$ they are subdominant in the neutrino matrix for the reason just showcased,
so eq(\ref{eq:Y_nu matrix}) is essentially unchanged by the contribution from eq(\ref{eq:Y_P2_P123}).

\subsection{Majorana masses\label{sub:Majorana}}

The heavy right-handed neutrino Majorana mass matrix has the largest contribution
from the operator of eq(\ref{eq:M_th_th}) giving the $(3,3)$ component 
\begin{equation}
\left( M_{N_{R}}\right) _{33}\simeq M_{3}\simeq \frac{\left\langle \bar{%
\theta}\right\rangle ^{2}}{M_{\nu _{R}}}  \label{eq:M3}
\end{equation}

The terms of eqs (\ref{eq:M_P23_P23}) and (\ref{eq:M_P123_P123}) give the
Majorana mass matrix of the form 
\begin{equation}
M_{N_{R}}\simeq M_{3}\left( 
\begin{array}{ccc}
\lambda _{1}\left( \frac{\epsilon _{\nu _{R}}}{\epsilon _{d}}\right)
^{4}\epsilon _{d}^{5} & \lambda _{1}\left( \frac{\epsilon _{\nu _{R}}}{%
\epsilon _{d}}\right) ^{4}\epsilon _{d}^{5} & \lambda _{3}\left( \frac{%
\epsilon _{\nu _{R}}}{\epsilon _{d}}\right) ^{4}\epsilon _{d}^{5} \\ 
\lambda _{1}\left( \frac{\epsilon _{\nu _{R}}}{\epsilon _{d}}\right)
^{4}\epsilon _{d}^{5} & \lambda _{2}\left( \frac{\epsilon _{\nu _{R}}}{%
\epsilon _{d}}\right) ^{4}\epsilon _{d}^{4} & \lambda _{4}\left( \frac{%
\epsilon _{\nu _{R}}}{\epsilon _{d}}\right) ^{4}\epsilon _{d}^{4} \\ 
\lambda _{3}\left( \frac{\epsilon _{\nu _{R}}}{\epsilon _{d}}\right)
^{4}\epsilon _{d}^{5} & \lambda _{4}\left( \frac{\epsilon _{\nu _{R}}}{%
\epsilon _{d}}\right) ^{4}\epsilon _{d}^{4} & 1%
\end{array}%
\right)  \label{eq:M matrix}
\end{equation}%
where the $\lambda _{i}$ are the $O(1)$ factors coming from the couplings
associated with the different operators of eqs (\ref{eq:M_P23_P23}) and (\ref%
{eq:M_P123_P123}).

\section{Phenomenological implications \label{sec:Phenomenology}}

It is now straightforward to determine the masses and mixing angles in the
theory. By construction the form of the up and down quark masses is in
agreement with the phenomenological fit of eq(\ref{eq:Yu}) and eq(\ref{eq:Yd}%
). If we further have $g_{\odot }=g_{@},$ giving a symmetric mass structure,
then the $(1,1)$ texture zero gives the successful GST
relation \cite{GST} relating the light quark masses and CP violating angle
to the mixing angle in the $(1,2)$ sector. A symmetric form for the mass
matrix is to be expected from the underlying $SO(10)$ and we will assume
this is the case here.

As discussed above, the charged lepton mass matrix gives the
phenomenologically successful relations $m_{b}\simeq m_{\tau }$ and $m_{\mu
}\simeq 3m_{s}$ at the unification scale. Moreover, the $(1,1)$ texture zero
implies that $Det[Y_{e}]\simeq Det[Y_{d}]$ so that $m_{e}\simeq m_{d}/3$ at the unification scale,
again in excellent agreement with experiment once one includes the radiative
corrections to the masses. The contribution to the mixing angles in the
lepton sector is given by 
\begin{eqnarray}
\theta _{12}^{l} &\simeq &\sqrt{\frac{m_{e}}{m_{\mu }}} \\
\theta _{23}^{l} &\simeq &\frac{m_{\mu }}{m_{\tau }}  \notag \\
\theta _{13}^{l} &\simeq &\frac{\sqrt{m_{e}m_{\mu }}}{m_{\tau }}  \notag
\end{eqnarray}

Finally we determine the neutrino masses and mixing angles. The Majorana
mass matrix has mass ratios given by 
\begin{eqnarray}
\frac{M_{1}}{M_{3}} &\simeq&\left( \frac{\epsilon _{\nu _{R}}}{\epsilon _{d}}%
\right) ^{4}\epsilon _{d}^{5} \\
\frac{M_{2}}{M_{3}} &\simeq&\left( \frac{\epsilon _{\nu _{R}}}{\epsilon _{d}}%
\right) ^{4}\epsilon _{d}^{4}  \notag
\end{eqnarray}%
Due to the large hierarchy in the Majorana mass matrix between $M_{1}$, $%
M_{2}$ and $M_{3}$, the contribution to the light neutrino masses from the
exchange of the third (heavy) right-handed neutrino is negligible. This is
despite the fact that the dominant Yukawa couplings in the Dirac mass matrix
are to the third right-handed neutrino, and realises the stategy discussed
in the introduction to explain the mismatch in the family symmetry breaking
patterns in the charged fermions and neutrino sector.

The light neutrino masses are given by 
\begin{equation}
m_{@}=\frac{\epsilon _{\nu_{L} }^{4}\epsilon _{d}^{2}h^{2}}{M_{1}}
\end{equation}

\begin{equation}
m_{\odot }=\frac{\epsilon _{\nu _{L}}^{4}\epsilon _{d}^{2}h^{2}}{M_{2}}
\label{eq:mn2}
\end{equation}%
\begin{equation}
m_{\nu _{3}}=\frac{\left( \frac{\epsilon _{\nu _{L}}}{\epsilon _{d}}\right)
^{4}h^{2}}{M_{3}}  \label{eq:mn3}
\end{equation}%
where $h$ is the doublet $H$ Higgs vev generating the up sector masses, and
where we have absorbed the $O(1)$ couplings in a redefinition of the
Majorana masses. Up to these $O(1)$ terms the ratio between the neutrino
masses is given by 
\begin{eqnarray}
\frac{m_{\odot }}{m_{@}} &=&\epsilon _{d} \\
\frac{m_{\nu _{3}}}{m_{\odot }} &=&\left( \frac{\epsilon _{\nu _{R}}}{%
\epsilon _{d}}\right) ^{4}\epsilon _{d}^{-2}\ll 1  \notag
\end{eqnarray}%

With our hierarchical mass structure the observed mass squared differences
relevant for atmospheric and solar oscillations are approximately given by $%
m_{@}^{2}$ and $m_{\odot }^{2}$ respectively. Up to the $O(1)$ coefficients, $%
m_{@}=$ $\epsilon _{d}\left( \frac{\epsilon _{\nu _{L}}}{\epsilon _{\nu _{R}}%
}\right) ^{4}\frac{1}{M_{3}}$, and a fit to atmospheric oscillation is
readily obtained by a choice of $\left( \frac{\epsilon _{\nu _{L}}}{\epsilon
_{\nu _{R}}}\right) ^{4}\frac{1}{M_{3}}$. With this normalisation the solar
oscillations have mass squared difference given by $m_{\odot }^{2}=\epsilon
_{d}^{2}m_{@}^{2}$. With the expansion parameter given in eq(\ref{eq:u,d eps}%
) from the down and charged lepton mass hierarchy fits and choosing
the $O(1)$ coefficients to be unity we obtain excellent agreement with the
magnitude found in solar neutrino oscillation.

The mixing angles are also readily obtained. To understand the results it is
convenient first to neglect the off-diagonal terms in the Majorana mass
matrix. The dominant exchange term in the see-saw mechanism is $\nu _{1}^{c}$%
. From eqs(\ref{Y33}) to (\ref{eq:Y_P23_P123}) we see that $\nu _{1}^{c}$
only couples via eq(\ref{eq:Y_P23_P123}) to the combination $\bar{\phi}%
_{23}^{i}\psi _{i}\varpropto (\nu _{\mu }-\nu _{\tau }).$ As a result the
most massive neutrino is close to bi-maximally mixed. The exchange of $\nu
_{2}^{c}$ is responsible for generating the next most massive neutrino. From
eqs(\ref{Y33}) to (\ref{eq:Y_P23_P123}) we see that it couples by both eq(%
\ref{eq:Y_P23_P123}) and eq(\ref{eq:Y_P123_P23}) to the combination $(\nu
_{\mu }-\nu _{\tau })+(\nu _{e}+\nu _{\mu }+\nu _{\tau }).$ Diagonalising
the masses the effect of this term is to introduce mixing at $O(\frac{%
m_{\odot }}{m_{@}})$ in the most massive state between the combinations $%
(\nu _{\mu }-\nu _{\tau })$ and $(\nu _{e}+\nu _{\mu }+\nu _{\tau }).$
However we have not yet introduced the effect of the off-diagonal terms in
the Majorana mass matrix, notably the entries 12 and 21, which also
introduce such mixing. Doing so we find that, due to the underlying symmetry
of the theory, these mixing terms cancel between the two contributions
giving 
\begin{eqnarray*}
\sin ^{2}\theta _{12}^{\nu } &=&\frac{1}{3} \\
\sin ^{2}\theta _{23}^{\nu } &=&\frac{1}{2} \\
\sin ^{2}\theta _{13}^{\nu } &=&0
\end{eqnarray*}%
Thus we get pure bi-tri-maximal mixing in the neutrino sector together with $%
\theta _{13}^{\nu }=0.$ Note that the underlying $SU(3)_{f}$ family
symmetry is responsible for these predictions.

Finally we should combine the contributions from the charged lepton sector
to obtain the PMNS angles. We should stress that the actual value of the corrections arising from the charged leptons depends on the value of the CP violating phase of the lepton sector, as shown in \cite{Rode}. Our model doesn't allow us to predict this phase independently (it originates from unknown phases of the fields involved in the vacuum alignment), but one can make a prediction of its value from the measured angles. Considering the cases that give us the largest deviations, we obtain a range of possible values for the angles given by

\begin{eqnarray*}
\sin \theta _{12} &\simeq &\frac{1}{\sqrt{3}}\left( 1\pm \theta
_{12}^{l}\right) \\
\sin \theta _{23} &\simeq &\frac{1}{\sqrt{2}}\left( 1\pm \theta
_{23}^{l}\right) \\
\sin \theta _{13} &\simeq &\frac{1}{\sqrt{2}}\theta _{12}^{l}
\end{eqnarray*}%
thus we have

\begin{eqnarray*}
\sin ^{2}\theta _{12} &\simeq &\frac{1}{3}\pm _{0.045}^{0.047} \\
\sin ^{2}\theta _{23} &\simeq &\frac{1}{2}\pm _{0.058}^{0.061} \\
\sin ^{2}\theta _{13} &\simeq &0.024
\end{eqnarray*}%
in good agreement with the experimentally measured values.

\section{Summary and conclusions\label{sec:Conclusion}}

In this paper we have shown how near bi-tri-maximal mixing in the lepton
sector arises from a spontaneously broken $SU(3)_{f}$ family symmetry
through vacuum alignment. The model constructed has a phenomenologically
acceptable pattern of quark and lepton masses. It generates the successful
GST relation between the mixing angles and masses of the first two
generations and the Georgi-Jarlskog relations between down quarks and
charged leptons. The neutrino sector generates precise bi-tri-maximal mixing
and a zero value for $\theta _{13}$. The charged lepton sector generates
small corrections to bi-tri-maximal mixing and a non-zero value for  $\theta
_{13}$. The see-saw plays a very important role in explaining the difference
between quark and lepton mixing angles. This is because the large family
symmetry breaking along the $3$ direction, which dominates the quark and
charged lepton masses, is suppressed in the light neutrino sector because
of the heavy right-handed neutrino mass in the $3$ direction that appears in the
see-saw mechanism. The model  has a fairly complicated multiplet content,
particularly in the symmetry breaking sector. However it represents only one
of a large class of such models and it may be hoped that more elegant
versions exist. In any case our example demonstrates the existence of a
phenomenologically satisfactory model in this general class.

\section*{Appendix - Vacuum Alignment}

As discussed above the alignment of vacuum expectation values is crucial to
the generation of bi-tri-maximal mixing. In this Appendix we consider the
minimisation of the scalar potential and show that with the terms in the
superpotential allowed by the symmetries of the theory the desired vacuum
alignment does indeed occur.

The field content and their symmetry properties are given in Table \ref%
{Ta:Table 1}.

We start with the $\theta _{i}$ and $\bar{\theta}^{i}$ fields. Their
soft masses $m_{\theta}$ and $m_{\bar{\theta}}$ have radiative
corrections coming from gauge and Yukawa related couplings. The gauge
couplings contribute positively to their the soft mass squared while the
Yukawa couplings contribute negatively. If the latter dominates their mass
squared can be driven negative, triggering radiative breaking and driving
nonzero vevs for the fields. Since  $\theta$ and $\bar{\theta }$ are in conjugate representations there is a  a $D-$flat direction with
equal vevs for  $\theta$ and $\bar{\theta }$. Assuming this
direction is also $F-$flat (this depends on the structure of the massive
sector of the theory which is not specified here) the scale of the vevs is
close to the point $\Lambda$, at which $m_{\theta}^{2}(\Lambda )+m_{%
\bar{\theta}}^{2}(\Lambda )=0$. Without loss of generality the
basis is chosen such that the breaking in the $SU(3)_{f}$ direction is
aligned along the $3$rd family direction.

Turning now to the $\phi _{3}$ and $\bar{\phi}_{3}$ fields we consider the
superpotential of the form 


\begin{equation*}
P_{3}= X_{3} \left(Tr[\bar{\phi}_{3}^{i}] \phi_{3_{i}} - M S_{3} \right)
\end{equation*}%

which is allowed by the symmetries of the theory (the trace is taken to yeld the $SU(3)_{f}\times
SU(2)_{R}$ invariant component of the first term).
M is a mass scale (it can arise from a singlet like $S_{3}$, as long as the charges are suitable). We assume that the field $S_{3}$ undergoes radiative breaking with vev $\mu _{3}$. Then
the $\phi _{3}$ and $\bar{\phi}_{3}$ vevs are triggered by the $F$-term $%
\left\vert F_{X_{3}}\right\vert ^{2}$.
These vevs develop
along the $D-$flat direction 
\begin{equation}
\left\langle \bar{\phi}_{3}\right\rangle =\left( 
\begin{array}{ccc}
0 & 0 & 1%
\end{array}%
\right) \otimes \left( 
\begin{array}{cc}
a_{u} & 0 \\ 
0 & a_{d}%
\end{array}%
\right) 
\end{equation}%
where the $SU(3)_{f}\times SU(2)_{R}$ structure is exhibited, and 
\begin{equation}
\left\langle \phi _{3}\right\rangle =\sqrt{a_{u}^{2}+a_{d}^{2}}\left( 
\begin{array}{c}
0 \\ 
0 \\ 
1%
\end{array}%
\right) 
\end{equation}%
Note that these vevs align along the same direction as $\theta$ and $%
\bar{\theta}$. The reason is that $\theta$ and $\bar{\theta}$ break $SU(3)_{f}$ to $SU(2)_{f}$ so that the stabilising
gauge radiative corrections to $\bar{\phi}_{3}$ act more on the 1st and 2nd
family elements than the third family element. As a result the third
components of $\bar{\phi}_{3}$ and $\phi _{3}$ are the ones that have the
smallest mass squared and so it is energetically favourable for the vevs to
develop along them.

The $\phi _{3}$ vev acquires a nonzero but small entry $\delta $, giving it the structure of eq(\ref{eq:P2 vev}). $\delta $ 
is fixed to preserve $D$-flatness (after the other vevs are triggered).

Consider now the adjoint field $\Sigma _{i}^{j}$, which we introduce for
alignment reasons. The symmetries of the theory allow the following
renormalisable terms in the superpotential%
\begin{equation}
P_{\Sigma }=\frac{\beta _{3}}{3}Tr\left( \Sigma ^{3}\right) +M\frac{\beta
_{2}}{2}Tr\left( \Sigma ^{2}\right) 
\end{equation}%
where $M$ is a mass scale in the effective theory associated with the
ultraviolet completion of the theory. It could be the breaking scale of the
underlying $SO(10)$ in a GUT implementation of the theory or could be
associated with the string scale in a string theory realisation. It
arises from the vev of a field with correct $R$ charge ($2/3$, same as that of $%
\Sigma $). This term induces a vev of the form%
\begin{equation}
\left\langle \Sigma \right\rangle =\left( 
\begin{array}{ccc}
a & 0 & 0 \\ 
0 & a & 0 \\ 
0 & 0 & -2a%
\end{array}%
\right) 
\end{equation}%
The relative alignment of $\left\langle \Sigma \right\rangle $ and $%
\left\langle \bar{\phi}_{3}\right\rangle $ follows because, with the vevs
aligned, both $\bar{\phi}_{3}$ and $\Sigma $ break $SU(3)_{f}$ $\ $to $%
SU(2)_{f}$ so that the stabilising radiative corrections to $\bar{\phi}_{3}$
act more on the 1st and 2nd family elements leaving the 3rd component
lightest and thus minimising the residual vacuum energy. 

Consider now the fields $\phi _{2}$, $\bar{\phi }_{23}$. In discussing the
cancellation of the $D-$terms we allow for the presence of additional fields 
$\bar{\phi}_{2}$, $\phi_{23}$ transforming in conjugate representations to $%
\phi_{2}$ and $\bar{\phi}_{23}$. The alignment of the vevs of these fields
is responsible for the difference of the mixing angles between the quark and
lepton sector. This alignment is driven by the following superpotential terms

\begin{equation}
P_{23}=Y_{2}Tr[\bar{\phi}_{3}^{i}]\phi _{2_{i}}+X_{23}\left( \bar{\phi}%
_{23}^{i}\phi _{3_{i}}\bar{\phi}_{23}^{j}\phi _{2_{j}}-\mu _{23}^{4}\right)
+Y_{23}\left( \bar{\phi}_{23}^{i}\phi _{23_{i}}\right) 
\end{equation}%
where the trace is over the $SU(2)_{R}$ indices and only the $SU(3)_{f}$
family indices are exhibited. The quantity $\mu _{23}$ is a mass scale, and
similarly to $\mu _{3}$ it can arise from an appropriate singlet $S_{23}$.
Following from this superpotential are $F-$term contributions to the scalar
potential which force the vevs of the form of eq(\ref{eq:P2 vev}), eq(\ref%
{eq:P23B vev}). The term $\left\vert F_{X_{23}}\right\vert ^{2}=$$%
\left\vert \bar{\phi}_{23}^{i}\phi _{3_{i}}\bar{\phi}_{23}^{j}\phi
_{2_{j}}-\mu _{23}^{4}\right\vert ^{2}$ forces both the vev of $\phi _{2}$
and of $\bar{\phi}_{23}$ to be nonzero. The term $\left\vert
F_{Y_{2}}\right\vert ^{2}=$$\left\vert \bar{\phi}_{3}^{i}\phi
_{2_{i}}\right\vert ^{2}$ forces the $\phi _{2}$ vev to be orthogonal to
that of $\bar{\phi}_{3}$. Without loss of generality we may choose a basis
in which it is aligned along the $2$ direction as in eq(\ref{eq:P2 vev})

\begin{equation}
\left\langle \phi _{2}\right\rangle =\left( 
\begin{array}{c}
0 \\ 
y \\ 
0%
\end{array}%
\right)
\end{equation}%
thus $\bar{\phi }_{23}$ must have vevs in both the $2$ and $3$ directions

\begin{equation}
\left\langle \bar{\phi}_{23}\right\rangle =\left( 
\begin{array}{ccc}
0 & b & b_{3}%
\end{array}%
\right)  \label{eq:P23B vev2}
\end{equation}

Consider now the pair $\phi_{123}$, $\bar{\phi}_{123}$. The alignment of
these will complete the vev alignment discussion. The relevant terms are

\begin{equation}
P_{123} = X_{123}\left( \bar{\phi}_{123}^{i}\phi _{123_{i}}-\mu
_{123}^{2}\right)  \label{eq:A_P123_X}
\end{equation}%
\begin{equation}
+Y_{123}\left( \bar{\phi}_{23}^{i}\phi _{123_{i}}\right)  \label{eq:A_P123_Y}
\end{equation}
\begin{equation}
+Z_{123}\left( \bar{\phi}_{123}^{i}\Sigma _{i}^{j}\phi _{123_{j}}\right)
\label{eq:A_P123_Z}
\end{equation}

The operators in eq(\ref{eq:A_P123_X}), eq(\ref{eq:A_P123_Y}) and eq(\ref%
{eq:A_P123_Z}) trigger and align the vevs of $\bar{\phi}_{123}$ eq(\ref%
{eq:P123B vev}) and $\phi _{123}$ eq(\ref{eq:P123 vev}) through the vevs of $%
\bar{\phi}_{23}$, $\Sigma $ and the mass scale $\mu _{123}$ (like with $\mu
_{23}$, this scale can be obtained though the vev of a singlet $S_{123}$).

Throughout the analysis the important effect of $D$-terms and soft
supersymmetry breaking mass terms $m_{i}$ must be taken into account, since
these terms play an important role in determining which vevs occur and in
vacuum alignment. We will analyse these conditions perturbatively in an
expansion involving small mass ratios, assuming the ordering $m_{i}\ll
\delta ,c,\bar{c}\ll b,b_{3},y\ll a_{u},$ $a_{d}$ which is the
phenomenologically viable range and which is readily obtained by a choice of
the free parameters of the theory. In this case minimisation of the
potential in leading order proceeds by setting the $D$ and $F$ terms to zero
and minimising the contribution to the potential coming from the soft terms.
Of course the true minimum corresponds to the case that the $D$ and $F$
terms are not zero but have a magnitude comparable to the contribution of
the soft terms. However this involves a nonleading change in the vevs found
setting the $D$ and $F$ terms to zero and so can be dropped in leading
order. In this way we can establish the local structure of the potential.

The interesting vacuum structure has the ($D$-flat and $F$-flat) form

\begin{equation}
\left\langle \bar{\phi}_{3}\right\rangle =\left( 
\begin{array}{ccc}
0 & 0 & 1%
\end{array}%
\right) \otimes \left( 
\begin{array}{cc}
a_{u} & 0 \\ 
0 & a_{d}%
\end{array}%
\right)
\label{eq:Pvev1}
\end{equation}

\begin{equation}
\left\langle \phi_{3}\right\rangle =\left( 
\begin{array}{c}
0 \\ 
\delta \\ 
\sqrt{a_{u}^{2}+a_{d}^{2}}%
\end{array}%
\right)
\end{equation}

\begin{equation}
\left\langle \bar{\phi}_{2}\right\rangle = \left( 
\begin{array}{ccc}
0 & \sqrt{y^{2}+\delta^{2}} & 0%
\end{array}%
\right)
\end{equation}

\begin{equation}
\left\langle {\phi}_{2}\right\rangle = \left( 
\begin{array}{c}
0 \\ 
y \\ 
0%
\end{array}%
\right)
\end{equation}

\begin{equation}
\left\langle \bar{\phi}_{23}\right\rangle =\left( 
\begin{array}{ccc}
0 & b & -b%
\end{array}%
\right)
\end{equation}

\begin{equation}
\left\langle {\phi }_{23}\right\rangle = \left( 
\begin{array}{c}
0 \\ 
b \\ 
b%
\end{array}%
\right) e^{i \beta}
\end{equation}

\begin{equation}
\left\langle \bar{\phi}_{123}\right\rangle =\left( 
\begin{array}{ccc}
\bar{c} & \bar{c} & \bar{c}%
\end{array}%
\right)
\end{equation}

\begin{equation}
\left\langle \phi_{123}\right\rangle =\left(%
\begin{array}{c}
\bar{c} \\ 
\bar{c} \\ 
\bar{c}%
\end{array}%
\right) e^{i \gamma}
\label{eq:Pvev2}
\end{equation}
where the overall phases are factored into the definitions of $b$ and $\bar{c%
}$ leaving only relative phases which are explicitly shown. These phases uniquely preserve the $F$-flatness in this configuration.

The relative phase of $\phi _{23}$ is connected with the relative
phase of $\bar{\phi}_{23}$ due to the \\ $\left\vert
F_{Y_{23}}\right\vert ^{2}=$$\left\vert \bar{\phi}_{23}^{i}\phi
_{23_{i}}\right\vert ^{2}$ orthogonality condition.

The $D$-term flatness
conditions constrain the relative phases of the $\phi _{123}$ and $\bar{\phi}%
_{123}$ components to be zero.

The relative phase of the $\bar{\phi}_{23}$
field components is $\pi$. This follows because the $F-$term $\left\vert
F_{Y_{123}}\right\vert ^{2}=$$\left\vert \bar{\phi}_{23}^{i}\phi
_{123_{i}}\right\vert ^{2}$ forces the respective vevs to be orthogonal.

To demonstrate that this is indeed a vacuum solution and prove the ratio of
the vevs, we must include the effect of the soft mass terms. We assume that
all the fields in discussion have positive mass squared.

Note that the vevs of $\bar{\phi}_{2},$ $\phi _{23},$ $\phi _{2}$ and $\bar{%
\phi}_{23}$ are triggered by minimising $\left\vert
F_{X_{23}}\right\vert ^{2}$, which requires $<ybb_{3}>=\mu _{23}^{4}/%
\sqrt{a_{u}^{2}+a_{d}^{2}}$. The mass term $\bar{m}_{23}^{2}\left\vert \bar{%
\phi}_{23}\right\vert ^{2}+m_{2}^{2}\left\vert \phi _{2}\right\vert
^{2}+m_{23}^{2}\left\vert \phi _{23}\right\vert ^{2}+\bar{m}%
_{2}^{2}\left\vert \bar{\phi}_{2}\right\vert ^{2}$ is then minimised by 
\begin{eqnarray*}
\left\vert b\right\vert  &=&\left\vert b_{3}\right\vert  \\
y^{2} &=&\frac{m_{23}^{2}+\bar{m}_{23}^{2}}{m_{2}^{2}+\bar{m}%
_{2}^{2}}b^{2}.
\end{eqnarray*}

Note that the equality of vevs in the 2 and 3 directions is a direct result of the
soft mass degeneracy of $\bar{\phi}_{23,2}^{2}$ and $\bar{\phi}_{23,3}^{2}$.
This follows from the underlying family symmetry $\bar{m}_{23}^{2}\left\vert 
\bar{\phi}_{23}\right\vert ^{2}=\bar{m}_{23}^{2}(\left\vert \bar{\phi}%
_{23,1}\right\vert ^{2}+\left\vert \bar{\phi}_{23,2}\right\vert
^{2}+\left\vert \bar{\phi}_{23,3}\right\vert ^{2})$ showing how the
required vacuum alignment is due to the family symmetry even though it is
spontaneously broken.

The remaining vacuum structure applies in a given region of soft mass
parameter space. The correct structure is obtained if $\left\vert \bar{m%
}_{23}\right\vert >\left\vert \bar{m}_{23}\right\vert$, $\left\vert \bar{m}%
_{123}\right\vert $, $\left\vert {m}_{123}\right\vert > \left\vert \bar{m}_{_{2}}\right\vert$. This ensures soft mass
minimisation won't allow $\bar{\phi}_{23}$ to be perturbed away from the vev
already determined. Then, as long as $\bar{\phi}_{2}$ soft mass is
light, minimisation of the soft terms favours the entries of $\phi _{123}
$ and $\bar{\phi}_{123}$ to be of equal magnitude (some were already
constrained to be so from $F$-flatness). The $D$-term highly
constrains the phases, allowing only overall phase ambiguities as shown in eqs(\ref{eq:Pvev1}) to (\ref{eq:Pvev2}). 

There is a longstanding problem associated with gauged family symmetry due
to the fact that the $D-$terms typically are non-vanishing and contribute to
the soft masses of the quarks and leptons in a family dependent way. This is
potentially disastrous as such masses can lead to unacceptably large
flavour changing neutral currents. Following Murayama \cite{murayama} one
sees that the effect is proportional to the difference in the mass squared
of the two fields developing large vevs along the $D-$flat direction. As a
result the effect can be suppressed if these masses are closely degenerate.
In \cite{SU3} we showed how this could naturally result in an $SU(3)_{f}$
model and the same structure readily emerges here. This is not the only way
the $D-$term contribution may be negligibly small. A specific example
follows when the SUSY\ breaking mediator mass is less than the family
breaking scale because the radiative graphs generating the dangerous soft
masses are suppressed by the ratio of the two scales. This will be the case
in this model for gauge mediated supersymmetry breaking. A more extensive
discussion of the suppression of $D-$term soft mass contributions will
appear shortly.

\begin{center}
\textbf{Acknowledgement}
\end{center}

The work of I. de Medeiros Varzielas was supported by FCT under the grant \\
SFRH/BD/12218/2003. This work was partially supported by the EC 6th
Framework Programme MRTN-CT-2004-503369

\end{document}